\documentstyle[12pt,aasms4]{article}

\begin{document}



\def \etal { et~al.\ }         
\def \m {M$_\odot$}          
\def \mm {M_\odot}           
\def \degrees{$^\circ$}



\lefthead{J. Stein, Z. Barkat and J.C. Wheeler}
\righthead{The Convective Urca Process}

\title{\LARGE{{\bf The Role of Kinetic Energy Flux in the 
Convective Urca Process}}}

\author{{\Large J. Stein$^{1,3}$, Z. Barkat$^{1,4}$ and J.C. Wheeler$^{2,5}$}}

\centerline{Submitted to Astrophysical Journal, September 1998}
\altaffiltext{1}{Racah Institute of Physics, Hebrew University of Jerusalem,
                Jerusalem, Israel}
\altaffiltext{2}{Department of Astronomy, University of Texas at Austin,
                 RLM 15.308, Austin, TX 78712-1083}
\altaffiltext{3}{E-mail: yossi@frodo.fiz.huji.ac.il}
\altaffiltext{4}{E-mail: zalman@saba.fiz.huji.ac.il}
\altaffiltext{5}{E-mail: wheel@astro.as.utexas.edu}

\bigskip
\begin{abstract}

The previous analysis of the convective Urca neutrino loss process 
in degenerate, convective, quasi-static, carbon-burning cores 
by Barkat \& Wheeler omitted specific consideration of the role 
of the kinetic energy flux.  
The arguments of Barkat \& Wheeler that steady-state composition 
gradients exist are correct, but chemical equilibrium does not
result in net cooling.  Barkat \& Wheeler included a ``work" 
term that effectively removed energy from the total energy budget that
could only have come from the kinetic energy, which must remain positive. 
Consideration of the kinetic energy in 
the thermodynamics of the convective Urca process shows that the 
convective Urca neutrinos reduce the {\it rate} of increase of entropy that 
would otherwise be associated with the input of nuclear energy and 
slow down the convective current, but, unlike the ``thermal" Urca
process do not reduce the entropy or temperature.

{\it Subject Headings}: Physical processes: convection $-$ hydrodynamics 
$-$ nuclear reactions $-$ Stars: interiors $-$ supernovae 

\end{abstract}

\section{Introduction}

The convective Urca process has been discussed for over two decades
in the literature and is still not satisfactorily understood or
resolved.  It should be just a matter of proper bookkeeping of
the thermodynamic variables, but the problem has proved complex
and subtle enough that even the sign of the effect is still debated.

The essence of the convective Urca process was first worked out by
Paczy\'{n}ski (1972) for which the convective circulation driven by
carbon burning in degenerate white dwarfs will yield first electron
capture and then $\beta$-decay of susceptible nuclei.  This will yield
no net change in composition, but a loss of neutrinos (or antineutrinos)
at each step of the cycle along with their attendant energy.  
Paczy\'{n}ski thus argued that this cycle would catalyze an energy loss
that would cool the star and postpone dynamical runaway.  
This was contested by Bruenn (1973) who pointed out that, microscopically,
each weak interaction added heat to the system, despite the loss of
the neutrino.  As electrons were captured below the Fermi sea,
another electron dropped from the Fermi surface to fill
the ``hole,''  resulting in heat,
or $\beta$-decay deposited electrons with excess thermal
energy above the Fermi sea on the other half of the cycle.  It is
important to determine the effect of the convective Urca process
because it has direct observational implications for Type Ia supernovae.
If a carbon/oxygen white dwarf of the Chandrasekhar mass undergoes 
thermonuclear runaway at the density at which carbon ignites, then
the density is relatively low and subsequent electron capture and 
neutronizing reactions in the explosion are minimized.  If, on
the other hand, the convective Urca process substantially postpones
the density of dynamical runaway to higher values, the attendant
neutronization in the thermonuclear combustion can lead to
unacceptably high neutron enrichement of the ejecta.

The physics of the convective Urca process was most recently analyzed
in detail by Barkat \& Wheeler (1990; hereafter BW), who argued that 
convective currents of composition play a critical role in the quasi-steady
state thermodynamics of the convective Urca process, so that cooling
terms associated with the convective flow exactly cancel the
microscopic heating terms described by Bruenn and lead to net
cooling. Their results seemed to be consistent with the results of the
careful numerical work of Iben (1978a,b, 1982) including a potential
understanding of the cooling instability of a convective core 
containing an Urca shell found by Iben.  In BW the instability resulted 
from the interaction of the convective core with the surrounding 
regions cooled by standard, plasma neutrino processes.

The conclusions of BW were called
into question by Mochkovitch (1996).   Mochkovitch presented a
general thermodynamic analysis and concluded that the convective
Urca process must heat the star.  The conclusions of Mochkovitch
were presented in a somewhat misleading way by implying that a
paper by Lazareff (1975) was correct in its analysis and conclusion 
that the rate of change of entropy, ds/dt, must be positive.  
The conclusion of Lazareff may be true, but his argument was
based on a very formal criterion that can be violated in principle,
as argued by BW.  The conclusion of Mochkovitch may
also be correct, but his analysis did not specifically reveal the
flaw in the physical arguments of BW.

In this paper we examine the issue yet again.  We conclude, in agreement
with Mochkovitch, that while the neutrinos associated with the
convective Urca process carry away energy, the entropy,
and the temperature, cannot decline.  Rather, the convective Urca
neutrino losses slow the convective currents and reduce the rate of 
increase of entropy associated with the input of nuclear energy. In \S 2, we 
make the case for the crucial role of the kinetic energy carried by the
convective current.  In the appendix we argue that proper treatment of the role
of kinetic energy was the physical component missing in the analysis of BW.
 BW included a ``work" term that
effectively removed energy from the total energy budget that
could only have come from the kinetic energy, which must
remain positive.   The loss rate by the Urca process cannot
exceed the rate of generation of kinetic energy.
The role of buoyancy and mixing in creating and limiting the
kinetic energy are outlined in \S 3 and \S 4 presents our conclusions.  

\section{Kinetic Energy of the Convective Current}

The average radial velocity of the matter in a white dwarf with central
carbon burning is negligible in the quasi-static phase because
for every upwelling current there is a downflow.  This does not
mean that the average speed of the convective current and the
kinetic energy associated with it are negligible.  The role of
this kinetic motion can be substantial even if the instantaneous
value of the kinetic energy of the flow is small in terms of
the total energy budget. 
As will be shown in \S 3, a significant portion of the nuclear energy
released is deposited in the form of a buoyancy potential energy that is 
subsequently converted into kinetic energy that is, in turn, dissipated
into thermal energy and Urca losses.  
The convective kinetic energy current is a key 
ingredient in turning the nuclear energy input that is produced in
a relatively confined volume of the core into thermal energy
spread throughout the convective core.

In many considerations of convection, the kinetic energy flux is treated 
as part of the ``convective luminosity," because one does not care 
whether heat is carried as thermal or kinetic energy, as long as 
it is spread around evenly to keep the entropy nearly constant in
regions of efficient convection. 
This picture is correct as long as the convective current is a passive
carrier of energy and particles, i.e. as long as the current itself
does not have to do actual work to move the material around.
When the convective Urca process is operating, however, there is a net 
electron current from the (average) radius at which the electrons are 
deposited by $\beta$-decay, to the (average) radius at which
they are recaptured. The convective current must carry these electrons 
``uphill," from low chemical potential to high 
chemical potential (Couch \& Arnett 1975).
This work is performed specifically at the 
expense of the kinetic energy of the convective flow,
not, directly, the nuclear input.
If there were no Urca neutrino losses, the kinetic energy of the current
would ultimately be expended as heat as the kinetic energy 
is dissipated by turbulent mixing.
With the Urca neutrino processes, some of the work done goes into the
energy carried off by the neutrinos.  This energy is then not available
to heat the matter.  

Picture a schematic version of the convective core in which the
nuclear input is confined in a rather small volume.  The nuclear
input creates a superadiabatic temperature gradient that generates
convection throughout a significantly larger volume.  The convection
is assumed to reach an Urca shell, where the Fermi energy is
equal to the difference in rest mass of Mother and Daughter nuclei
(plus an electron) so $\beta$-decay and neutrino loss will occur, 
converting Mothers to Daughters.  
Subsequently, convection will carry Daughters inward and they will be 
converted back to Mothers with the production and loss of an anti-neutrino.
Consideration of the work done in this cycle and 
the kinetic energy of the convective flow
puts limits on the amount of energy that
can be carried out by the neutrinos:
\begin{itemize}
\item[a.] Only a fraction, $f_1$, of the nuclear energy 
input ($q=q_{nuc}-q_\nu$)
can be converted into kinetic energy.
\item[b.] Because of the dissipation associated with 
turbulent mixing, only a fraction, $f_2$, of the kinetic energy can
reach the Urca shell.  This fraction may depend on the nature of
the mixing, e.g. eddies or plumes.
\item[c.] Only a fraction, $f_3$, of the work done by 
the kinetic energy in lifting
the Urca nuclei against the gradient in Fermi energy 
is converted into neutrinos.  The remainder (about 25 percent, Bruenn(1973)) is
converted to heat, thus $f_3 \leq 3/4$.

It is then immediately clear that Urca neutrino losses {\it can not amount to
more than $ \sim \frac{3}{4} q$,} even if $f_1 = f_2 = 1$.

\end{itemize}

\noindent
We shall show in the next section that in the quasi-steady-state 
$f_1$ can be reasonably well estimated.  The fraction
$f_2$ is hard to estimate.

We argue that the net effect of the convective Urca 
process is twofold:

1. It {\it reduces the rate of increase} of the entropy due to
nuclear burning.   The convective Urca process includes 3 components:
$\beta$-decay/capture, mixing, and convection.  The first two
components increase the entropy and the third cannot decrease it.
The convective Urca process thus cannot reduce the entropy in an 
absolute sense, and likewise cannot reduce the temperature 
(for constant or increasing density), only the rate of increase of 
temperature.

2. It slows down the speed of convection beyond the Urca shell. 
Under the usual assumption that convective timescales are rapid 
compared to those of the $\beta$-processes, the rate of capture and 
decay of Urca nuclei depends only on the extent of the convective zone
on both sides of the Urca shell.  When the Urca shell is not too close 
to the center of the star, which is typically the case,
convection is stopped shortly beyond
the Urca shell. The heat created by the Urca process near the Urca shell
further tends to flatten the temperature gradient locally there
and hence to reduce the buoyancy within the convective core
and thus contributes to the slowing down of the convection.
The restriction of the size of the convective core by the Urca
process limits the entropy at the edge of the convective zone. Since
the expansion of the convective core is restricted, the core is prevented
from growing to the traditional limit where the (nearly constant)
entropy of the inner convective core is equal to the entropy of the 
outer stable layers with positive entropy gradient.  The likely result is that
a boundary region of negative entropy gradient will be established
between the fully convective core and the outer stable regions.

In the appendix we examine the role of the kinetic energy
by examining the neglect of its explicit treatment by BW.

\section{Buoyancy and Kinetic Energy.}

The Urca loss rates are limited by the kinetic energy in 
the convective flow.  If the losses were to become so
strong that the currents were slowed, the losses would be limited.
  A limit to the kinetic energy in the
convective flow is given by the rate of production of the
buoyancy potential energy by the thermonuclear reactions.
 Here we estimate the maximum 
rate at which buoyancy potential can be generated, giving an upper limit
to the rate of creation of the kinetic energy (i.e. $f_1$).

Consider a convective region at the center of the star, with constant 
entropy density gradient $\frac{\partial{s}}{\partial{M}}$. 
At the center of the convective zone there is a heat source $q_{nuc}$.
There is no other source or sink in the convective zone.
For simplicity, neglect conductive heat transport and
the ``normal" neutrino losses that
would limit the convective core if there were no Urca process.
We choose a time step $dt$, which is
large compared to the convection turnaround time, but small enough,
that $q_{nuc} dt$ is small.
Let us assume for the moment that the matter outside the initial
convective region does not move, and so all the heat produced
is contained in this initial convective region.
The heat created during the interval $dt$ is: 
\begin{equation}
dQ = (\int_{0}^{M_{cc}} q_{nuc} dM)dt,
\end{equation}
where $M_{cc}$ is the mass of the convective core.
The associated total entropy change of the convective region is then:
\begin{equation}
 dS = (\int_{0}^{M_{cc}} {q_{nuc}\over T} dM)\,dt = 1/T_q dQ,
\end{equation}
where $1/T_q$ is an average of $1/T$ over the region where $q_{nuc} > 0$:
\begin{equation}
 1/T_q \equiv (\int_{0}^{M_{cc}} {q_{nuc}\over T} dM)
      /(\int_{0}^{M_{cc}} q_{nuc} dM).
\end{equation}
To the extent that $q_{nuc}$ is highly centrally concentrated,
$T_q$ will be close to the central temperature of the core.

 Within the convective region, the mass that is heated by the nuclear
reactions at the center of the region is assumed to move adiabiatically
to rapidly mix its heat smoothly throughout the convective region
by sharing its entropy with the larger mass of the whole convective
region.  The resulting change in the specific entropy, 
ds = dS/M$_{cc}$, is then independent of the postion within the 
convective core.  We could view this process as comprising
two steps: coarse mixing by circulation that spreads the heated
mass evenly over the convective zone without changing the entropy,
then a fine local mixing, which changes the entropy. This change of entropy
in the second step is a second-order effect and can be ignored.
With these assumptions, the thermal energy of the convective zone is increased
by:
\begin{equation}
 dE = \int_0^{M_{cc}} (T ds) dM = {dS\over M_{cc}} \int_0^{M_{cc}} T dM
     = dS~T_{cc} = T_{cc}/T_q dQ,
\end{equation}
where $T_{cc}$ is an average of $T$ over the convective zone:
\begin{equation}
 T_{cc} \equiv (\int_{0}^{M_{cc}} T dM)/(\int_{0}^{M_{cc}} dM).
\end{equation}      

The buoyancy $dB$ is the difference between the energy $dQ$ released by
the reaction and the final change $dE$ in the thermal energy:
\begin{equation}
 dB = dQ-dE = dQ (1 -  T_{cc}/T_q),
\end{equation}
or:
\begin{equation}
 f_1 = 1 -  T_{cc}/T_q
\end{equation}

The value of $f_1$ can be computed straightforwardly for a 
convective core with no Urca processes.  The presence of the
Urca process that restricts the growth of the convective core
will generate a region with negative entropy just beyond
the Urca shell, as mentioned in \S 2.  
The effect of this boundary gradient on the fraction
of the input energy that is converted to kinetic energy is
difficult to estimate.  The effect of this region of negative 
entropy gradient is being explored with 2D numerical simulations.

\section{Conclusions}

We conclude here: a) In agreement with Mochkovitch (1996)
and in disagreement with the earlier work in BW,  that 
the convective Urca process can reduce the rate of
heating by nuclear reactions, but cannot result in a net
decrease in entropy, and hence in temperature for a
constant or increasing density;  b) The convective Urca process must
limit the expansion of the convective zone beyond the Urca shell.
The error in BW is traced back to ignoring the role of the kinetic energy
(See appendix).

An interesting precursor
to these conclusions is the work of Finzi \& Wolf (1968)
on the pulsational Urca process.  Finzi \& Wolf show
that a system makes a transition from a ``thermal Urca"
regime to a ``vibrational Urca" regime when the 
vibrational energy becomes large compared to kT.  The
rate of dissipation of the vibrational energy grows with the 
vibrational energy with part of the dissipated energy
going into neutrino losses and the remainder into thermal
energy. The dissipation of the vibrational energy thus drives
the pulsational Urca neutrino losses. The neutrino losses never, of course,
dissipate vibrational energy faster than it is being pumped in,
and the vibrational energy can only be reduced to zero without
cutting off the pulsational neutrino loss process.  In our
problem, the convective Urca losses grow as the convective
core extends further beyond the Urca shell. The limit established above on
Urca losses thus prevents the convection from extending far beyond
the Urca shell.

While we agree with the general conclusions of Mochkovitch (1996),
we feel that his simple picture of the convective core should,
perhaps, be taken with a note of caution.  He follows
the argument of Couch \& Arnett (1975) to write the work done
by convection as $w = \epsilon_F^{ec} - \epsilon_F^{em}$ and
then uses this equality to establish some interesting inequalities that
suggest that convective Urca losses cannot stabilize nuclear 
burning and that the convective heat engine needs a cold source,
the material at the boundary of convective core.  These conclusions
may be true, but the 
work done by convection must not only account for 
the difference in the Fermi energy at electron capture and emission,
but also the energy loss by dissipation of the convective currents
and thus $w > \epsilon_F^{ec} - \epsilon_F^{em}$.

Over the last couple of decades most work on modeling of the
progenitors of Type Ia supernovae has simply ignored the
effects of the convective Urca process.  
The conclusion we have reached, after similar decades of pondering, is that 
that was probably the right approach.

We are grateful for conversations with Robert Mochkovitch who 
started us thinking about this problem, yet again, and to
Steve Bruenn, Raph Hix, Icko Iben, Jr. and Alexei Khokhlov
for discussions of the topic.
This research was supported in part by NSF Grant AST-9528110,
NASA Grant NAG5-3930 and a grant from the Texas Advanced Research Program.  

\appendix

\section* {Appendix: The Barkat - Wheeler Paper}
BW do not refer directly to the convective kinetic energy.
They define the convective luminosity profile
$L_{cc}(M, t)$ which is assumed to adjust instantaneously to
the slowly varying nuclear heat source, q.  
The equations of \S 2 of BW are consistent with  
the interpretation that the kinetic energy is included in 
the ``internal energy'' carried by the convective flux
and hence that $L_{cc}$ represents the flux of both the
thermal and kinetic energy (note that the small contribution
from the diffusion of energy is ignored). The
kinetic energy can be created at a high rate by the nuclear
reactions and yet destroyed at a high rate by turbulent dissipation. 
The kinetic energy can hence be small at any given instant
in the quasi-static approximation and almost constant in time.

In addition to its conceptual and physical role in the conversion
of nuclear to thermal and neutrino energy, it is important to keep 
track of the kinetic energy separately even though it may be small, 
because {\it the kinetic energy must be positive}, and it is the
kinetic energy that drives the Urca process and that is converted 
into neutrinos and heat by the Urca process.
This allows an upper limit to be placed on the Urca losses.
The Urca losses can only remove energy at a rate that does not reduce the
kinetic energy to zero, otherwise the convective Urca process
will shut down.

Here we will argue that the formulation of BW was essentially
correct with the interpretation that the internal energy
includes the kinetic energy.  The problem arises in BW
because the Urca losses come from the kinetic energy and
the kinetic energy is positive.  This effectively restricts
the Urca losses.  If the kinetic energy is treated as an
indistinguishable part of the total internal ``thermal" energy,
then this restriction is not manifest.   

In particular, BW
correctly argue that the Urca convective core must be close
to chemical equilibrium with gradients in the composition
of the Urca-active components that lead to steady-state
currents.  BW err in the manner in which they invoke 
chemical equilibrium in the entropy equation.  BW include a 
term that implies that the current of electrons does work
against the full electron chemical potential gradient
and subtract that work, the critical ``cooling term," from 
the ``heating terms" in the entropy equation. 
We stress the crucial point that this subtraction was done in 
the entropy equation, not in the energy equation.  
The treatment in BW in essence says that the
current of electrons brings electrons with a {\it negative} entropy
equivalent to -$|\epsilon_F$-$\epsilon_{th}|$/T 
into the Urca shell (and carries out electrons with
{\it positive} entropy $|\epsilon_{th}$-$\epsilon_F|$/T 
on the other side of the Urca shell), where $\epsilon_F$ is
electron Fermi energy, and $\epsilon_{th}$ is the threshold energy for
electron capture. BW added that negative entropy to
the entropy in a given mass element.  Instead, the entropy
brought in by the current of Urca-active elements is small, 
only that represented by the thermal component of the electron energy,
and it is positive.  
Since $\epsilon_F$-$\epsilon_{th}$ $>>$ kT except directly at
the Urca shell, the effect of the term invoked by BW is 
to drastically exaggerate the loss of entropy.

BW correctly write the expression for the loss of entropy
due to neutrino processes in the absence of convection
in their equation 8:
\begin{equation}
T\frac{ds}{dt} = q_{nuc} - q_{\nu} - \frac{\partial L_{diff}}{\partial M}
 - \epsilon_{\nu}\mid\frac{dN_e}{dt}\mid - \sum_{i}\mu_i\frac{dN_i}{dt},
\end{equation}
where $q_{nuc}$ is the nuclear input, $q_{\nu}$ is the standard
(plasmon) neutrino loss, $\frac{\partial L_{diff}}{\partial M}$  is
the (negligible) energy carried by radiation diffusion, and
$\epsilon_{\nu}|\frac{dN_e}{dt}|$ represents the non-convective
contribution of the thermal Urca processes.  Since 
$\sum_{i}\mu_i\frac{dN_i}{dt}$ = 
$-|(\epsilon_F - \epsilon_{th})||\frac{dN_e}{dt}|$,
Eq. 1 can be written as:
\begin{equation}
T\frac{ds}{dt} = q_{nuc} - q_{\nu} - \frac{\partial L_{diff}}{\partial M}
 + (|\epsilon_F -\epsilon_{th}| - \epsilon_{\nu})\mid\frac{dN_e}{dt}\mid,
\end{equation}  
where the last term summarizes the net Urca heating.  BW define
$\epsilon_+ = |\epsilon_F -\epsilon_{th}|$ and $\epsilon_h$ = 
$\epsilon_+ - \epsilon_{\nu})$ such that the final heating term in 
Eq. 2 is $\epsilon_h|dN_e/dt|$.

The correct expression for the entropy in the presence of
a convective flow and associated neutrino losses is:
\begin{equation}
T\frac{\partial{s}}{\partial{t}} 
  = T\frac{ds}{dt} - \frac{\partial L_{cc}}{\partial M} ,
\end{equation}      
where the local contribution to the change in entropy Tds/dt
is given by Eq. 2, and the second term
on the right hand side represents the contribution
due to the currents, i.e., the convective luminosity.  
The details of how one should write the convective luminosity
are difficult to define and implement
in practice and may not be important, given our basic
conclusion that, at best, the Urca process can only
limit the rate of growth of entropy.

In their attempt to include the effects of Urca convection
on the entropy balance, BW write their equation 29:
\begin{equation}
T\frac{\partial{s}}{\partial{t}} 
    = q_{nuc} - q_{\nu} - \frac{\partial L_{diff}}{\partial M}
 - \epsilon_{\nu}\mid\frac{dN_e}{dt}\mid - \frac{\partial L_{cc}}{\partial M}  
 - \sum_{i} \mu_i\frac{\partial{N_i}}{\partial{t}}.
\end{equation}  
Note that there is a typo in BW equation 29, the entropy derivative
is with respect to time, not temperature, and that the
term q is not clearly defined.  The term ``q" as used in BW
Eq. 29 is not the q
of BW equation 5, but rather the first four terms of Eq. 4 as we
have written them explicitly here.
The error in BW is in the term 
$-\sum_{i} \mu_i\partial{N_i}/\partial{t}$.  BW break this sum into two terms,
the local Lagrangian term $-\sum_{i}\mu_i dN_i/dt$ as included in Eq. 2 here,
and a term due to the current of electrons, 
$\sum_{i} \mu_i \partial{L_{N_i}}/\partial{M}$.  
The problem is with the latter term.  The current of electrons
and Urca-active nuclei cannot carry away the implied quantity
of entropy.

The meaning of the term $\sum_{i} \mu_i \partial{L_{N_i}}/\partial{M}$
becomes more clear when we integrate it over the whole convective region.
Integrating by parts, and remembering that $L_{N_i}$ vanishes on the
boundaries, we get:
\begin{equation}
   -\int_0^{M_{cc}} {\sum_{i} {L_{N_i} \partial{\mu_i}/\partial{M}}} =
   -\int_0^{M_{cc}} {\sum_{i} {L_e \partial{\epsilon_f}/\partial{M}}},
\end{equation}
where we have neglected the chemical potential of the ions.
The absolute value of the last expression is the work done
by the current in pushing the extra electrons ``uphill."
Therefore the expression in Eq. 5  represents the loss of 
kinetic energy associated with this work and not the change in entropy.
This term should {\it not} have been included in the entropy
equation Eq. 29 of BW.  For illustration, we have included the
same erroneous term here in Eq. 4, but it is incorrect
for the same reason.  BW Eq. 30 is formally correct, representing
the translation from Lagrangian to Eulerian coordinates, but not
applicable to Eq. 29, since the last term in that equation and
in Eq. 4 as written here should be the Lagrangian component only
(as given here in Eq. 1), not the Eulerian component.

BW employ the argument that by chemical equilibrium a suitable
average over the convective core
must give $\sum_{i} \mu_i\frac{\partial{N_i}}{\partial{t}} = 0$.
They thus effectively argue that the last term in Eq. 4 (or
their Eq. 29) is zero and that the remaining terms result in
net cooling. To see this in a bit more detail, 
this expression can be decomposed as 
$\sum_{i} \mu_i\frac{\partial{N_i}}{\partial{t}} =
\sum_{i} \mu_i\frac{dN_i}{dt} - \sum_{i} \mu_i\frac{\partial{L_{N_i}}}
{\partial{M}} = 0$.
Note from the discussion following Eq. 1 that $\sum_{i}\mu_i\frac{dN_i}{dt}$ = 
$-\epsilon_+|\frac{dN_e}{dt}|$.  The application of chemical 
equilibrium thus also gives $\sum_{i} \mu_i\frac{\partial{L_{N_i}}}
{\partial{M}}$ = $-\epsilon_+|\frac{dN_e}{dt}|$.  The latter term
is thus a putative cooling term due to 
currents that offsets the heating term due to local Urca heating.
Since this ``cooling" term should not be present in Eq. 4, we conclude that
chemical equilibrium is a reasonable assumption for this system,
but that it does not play any direct role in the analysis of
the thermodynamics of the convective Urca process.  Given the
erroneous introduction of the ``cooling" term in their Eq. 29, the 
subsequent discussion in BW (through Eq. 37 ff) is irrelevant. We note
that Eq. 37 as presented in BW, even if corrected, is not a 
good way to represent the growth of entropy in time.  The analog
would be an integral of Eq. 1 over the mass of the convective core, but such
a procedure would neglect the strong feedback of the Urca process
on the convective current. 
Inclusion of the term in Eq. 5 in the energy equation, 
as done by Iben (1978a; Eq. 8), is allowed. Care must be taken,
however, to include its affects only on the kinetic energy,
which must always be non-negative.  

Much of the past work on the convective Urca process has
assumed that the size of the convective core is defined
by the condition that the convective Urca losses (plus
minor standard neutrino losses) be equal to the rate
of nuclear input.  Since the convective Urca process cannot
actually reduce the entropy of the convective core,
this definition of the convective core is invalid. The
size of the convective core must instead be set by
the normal processes of the entropy profile plus
some overshoot.  The convective Urca process may affect
the size of the core by slowing the speed of convective 
currents when the boundary of the core exceeds the radius
of an Urca shell.  More specifically, we note that
the core stability analysis of BW is invalid because
it depended specifically on equating the Urca losses with
nuclear input.

References

Barkat, Z. \& Wheeler, J. C. 1990, ApJ, 355, 602.

Bruenn, S. W. 1973, ApJ, 183, L125.

Couch, R. G. \& Arnett, W. D. 1975, ApJ, 196, 791

Finzi, A. \& Wolf, R. A. 1968, ApJ 153, 835

Iben, I. I., Jr. 1978a, ApJ, 219, 213

Iben, I. I., Jr. 1978b, ApJ, 226, 996

Iben, I. I., Jr. 1982, ApJ, 253, 248

Mochkovitch, R. 1996, A\&A 311, 152

Paczy\'nski, B. E. 1972, ApLett, 11, 53

\end{document}